# Near band gap photoluminescence properties of hexagonal boron nitride

**Luc MUSEUR**

*Laboratoire de Physique des Lasers – LPL, CNRS UMR 7538, Institut Galilée, Université Paris 13, 93430 Villetaneuse, France*

**Andrei KANAEV**

*Laboratoire d'Ingénierie des Matériaux et des Hautes Pressions – LIMHP, CNRS, Institut Galilée, Université Paris 13, 93430 Villetaneuse, France*

# Abstract.

Near band-gap luminescence (hν≥5 eV) of hexagonal boron nitride has been studied by means of the time- and energy-resolved photoluminescence spectroscopy method. Two emissions have been observed at 5.5 eV and 5.3 eV. The high-energy emission at 5.5 eV is composed of fixed sub-bands assigned to bound excitons at 5.47 eV, 5.56 eV and 5.61 eV. The non-structured low-energy emission at 5.3 eV undergoes a large blue shift (up to 120 meV) with a linear slope $\Delta E_{lum} / \Delta E_{exc}$ <1 with increasing excitation energy $E_{exc}$. At $E_{exc} \geq 5.7$ eV, the band position is fixed and marks the transition from the Raman to the photoluminescence regime. We assign the 5.3 eV band to quasi donor-acceptor pair (q-DAP) states due to electrostatic band fluctuations induced by charged defects. The shift is explained by photo-induced neutralization of charged defect states. The absence of contribution to the q-DAP luminescence from exciton suggests the existence of a large exciton binding energy, which is qualitatively consistent with theoretical predictions.

# I. Introduction.

Graphite-like hexagonal boron nitride (hBN) is the most commonly used polymorph of boron nitride which also exits in cubic (diamond-like), rhombohedral, and wurzitic forms. This interest results from important properties of hBN such as a high electrical resistance, a high thermal conductivity, and an elevated melting point combined to an excellent resistance to the oxidation that make this chemically inert material interesting in a wide range of applications. Its optical properties are also of great interest because of the hBN band gap energy which is one of the largest in the III – nitrides family. This interest has been recently renewed following the observation of far-UV (215 nm) laser action in a single crystal under electronic excitation[1]. Moreover, under specific experimental conditions, hBN is known to form single- and multi-wall nanotubes. In contrast with carbon nanotubes, the band gap energy of hBN nanotubes is predicted to be nearly independent of their diameter and helicity[2]. These properties make hBN, either in bulk or nanotubes forms, a very promising material for optoelectronic applications in the UV range.

Despite many efforts, the electronic and optical properties of bulk hBN remain largely unknown. For example, there is still no agreement on the band gap value and its nature (direct or indirect transition). From an experimental point of view, the dispersion of the band-gap energy reported in literature, ranging from 4 eV to 7 eV, is ascribed to the low-purity of poly-dispersed microcrystalline of the commonly available samples. In the past, the luminescence properties of hBN have been studied by numerous groups by cathodoluminescence [1, 3-6] or photoluminescence [3, 7-12] spectroscopy. Depending on experimental conditions, two structured emission bands are reported around 4 eV and 5.5 eV. Their origin is still controversial and assignments provided in the past have to be revised in the view of recent band gap measurements[1] and calculations[13, 14].

Recently, thanks to cathodoluminescence and absorption measurements performed on a high-quality single crystal, Watanabe *et al.* have suggested the existence of a direct band-gap at 5.98 eV and assigned the strong luminescence observed at 215 nm (5.76 eV) to a Wannier exciton with a binding energy of 0.15 eV [1]. Conversely, very recent theoretical calculations of electronic structures and optical spectra disagree with these conclusions and predict indirect band-gap properties for hBN [13, 14]. Based on the so-called all- electron GW approximation, Arnaud *et al.* have obtained an indirect band gap transition in the range 5.5 and 6 eV [14] and have predicted a huge excitonic contribution to the optical spectra. The lowest excitonic state at 5.85 eV was predicted to be of Frenkel type with a strong binding energy of 0.72 eV referred to the hBN direct band-gap [14]. In view of these disagreements, a careful experimental analysis of BN electronic transitions is of particular interest.

Near band-gap optical properties of hBN is a key issue in the understanding of the electronic structure and energy-transfer mechanisms of this promising semiconductor. In the present communication, we report on detailed experimental studies conducted at low-temperature of the hBN near band-gap luminescence excited by synchrotron radiation (SR). Low-intensity SR enables the observation of excitonic and donor acceptor pair (DAP) transitions in conditions almost free of undesirable excitation saturation effects. A comparison is made with two kinds of samples: powder-like hBN (pd-hBN) and pyrolytic hBN (pBN). Based on spectral and time-resolved measurements, we provide evidence on the nature of the observed luminescence bands. We discuss also the binding energy of the observed exciton.

## II. Experiment.

Two kinds of hexagonal boron nitride samples were analyzed. The first sample (pd-hBN) was made from commercial hexagonal BN powders (Alfa 99.5%) compacted in a square pellet (8x8x1 mm$^3$) under a hydrostatic pressure of 0.6 GPa. The grit size of the hBN powder has been estimated by means of granulometry and transmission electron microscopy (JEM 100C JEOL). It ranged from 0.3 to 10 μm with an average particle size of 3.1 μm, corresponding to the maximum in the mass distribution curve. In our experiments, no specific results have been obtained with hBN samples heated at 800 K under vacuum for a period of 12 hours to avoid trace contaminants such as organic impurities or water. The second kind of samples was pyrolytic hBN (pBN) deposited on a Si substrate by CVD method through a gas-phase reaction between $BCl_3$ and $NH_3$ at 2300 K. This technique, in contrast to the traditional ceramic one, ensures considerably more pure samples free of carbon compounds. From a structural point of view, pBN is a layered material composed of small hexagonal crystallites (~ few nm) that are highly oriented in the direction perpendicular to their basal planes, but randomly oriented within the layer. Consequently, the density of stacking faults is higher in pBN than in hBN microcrystals composing the powder. On the other hand, due to the strong contact between the crystallites composing the in pBN, the specific surface is much lower in the pBN, as compared to the pd-hBN.

The luminescence properties of the samples were measured by means of the VUV synchrotron-radiation excitation (SR) from the DORIS storage ring at HASYLAB (DESY, Hamburg). The facility of the SUPERLUMI station used for experiments is described in details elsewhere[15]. Briefly, samples were cooled down to 8K and irradiated by monochromatized SR ($\Delta\lambda = 3.3 \text{\AA}$) under high vacuum (~$10^{-9}$ mbar). The measurements of luminescent spectra were carried out using a visible 0.275-m triple grating ARC monochromator equipped with a CCD detector or a photomultiplier operating in the photon-counting mode. The pulse structure of the synchrotron radiation (excitation pulses of 130 ps at 5 MHz repetition rate) allows time-resolved luminescence analysis at time-scale ≤200 ns with sub-nanosecond resolution. Luminescence excitation spectra were recorded within a time gate Δτ delayed to the SR excitation pulse. Two time-gates were used simultaneously: a fast one ($\Delta\tau_1 = 2-6 \text{ ns}$), and a medium one ($\Delta\tau_2 = 10-30 \text{ ns}$). Complementary luminescence decay curves were measured at fixed excitation- and luminescence-energy.

The recorded spectra were corrected for the primary monochromator reflectivity and SR current. The luminescence spectra were corrected for secondary monochromator reflectivity and CCD sensitivity.

## III. Results.

Two luminescence spectra of pd-hBN and pBN under 6-eV excitation are displayed in Figure 1. According to the excitation wavelength, several luminescence bands are observed. The luminescence spectra of the hBN compacted powder (pd-hBN) appears to be richer as compared to that of pBN. This underlines the importance of impurities in the photoluminescence process. Both spectra display broad bands in the range of 3.5 – 4 eV. The time-resolved luminescence analysis allows the assignment of these bands to the radiative recombination of a donor-acceptor pair (DAP). Moreover, in the case of the pd-hBN sample (Figure 1.b), the small modulation observed on the top of the band around 4 eV can be assigned to defect transitions involving impurity atoms. These issues will be discussed in a forthcoming article[16]. Here, we focus on the narrower emission bands situated at higher energies, between 5.3 and 5.5 eV.

At excitation energies of the pd-hBN sample above 5.5 eV, two intense luminescence bands appear at 5.3 eV and 5.5 eV (Figure 1.b). Their integral intensity is about one third of the broad emission at 4 eV. In contrast, the intensity of the 5.3 eV band observed in pBN sample (Figure 1.a) is very low (~4·10$^{-2}$). These two bands have been reported earlier in cathodoluminescence experiments using thin films of hBN [5], single crystal [1] or commercial powder [3]. Up to now, their precise nature is subject to debate. Radiative recombination of bound Wannier excitons [1] or Frenkel excitons [3] has been proposed. However, due the energy difference between the peaks, which coincides with the energy of the in-plane LO phonons in hBN (~199 meV), a phonon-assisted radiative recombination process has also been suggested [5]. In order to elucidate the nature of these bands, we have carried out complementary measurements with a higher spectral and time resolution.

Series of high-resolution photoluminescence spectra obtained in pd-hBN sample are presented in Figure 2. The higher energy 5.5-eV band is actually composed of three narrow sub-bands at 5.47 eV, 5.56 eV and 5.61 eV. These bands appear for excitation energy above 5.7 eV and their position are invariant to the excitation energy. Conversely, no sub-structure is detected for the lower energy band, which undergoes a strong energy shift when the excitation energy, $E_{ex}$, changes. This band appears around 5.18 eV for excitation at 5.5 eV. The intensity of this band increases with the excitation energy, with a concomitant blue shift of the peak maximum. With excitations above 5.7 eV, the spectral position remains fixed at 5.3 eV. The remarkable difference in the spectral behaviour of the 5.3 eV and 5.5 eV bands is a strong indication of their different origins. This conclusion is supported by the photoluminescence excitation spectra (PLE) measurements which are sensitive to the various type of energy transfer processes.

The excitation spectra of 5.3 and 5.47 eV luminescence in pd-hBN sample are presented in Figure 3. The 5.47 eV PLE spectrum is characterized by a steep rise at the energy threshold of $E_{exc}$ = 5.6 eV and a broad maximum at $E_{exc}$ = 5.81 eV. This last value is in reasonable agreement with that of the free exciton in hBN single crystal (5.76 eV) [1]. The 5.3 eV PLE spectrum looks qualitatively similar: a steep rise is observed at the energy threshold of $E_{exc}$ = 5.5 eV and the maximum is obtained at 5.7 eV. However, these values are somewhat lower than the corresponding ones in the 5.47 eV PLE spectrum. A noticeable feature of the 5.3 eV PLE spectrum is the exponential dependence on the excitation photon energy, $E_{exc}$. A least-square fit with $I_{lum} \propto \exp(E_{exc}/E_0)$ yields $E_0$ = 41 meV and is displayed by a dashed line in Figure 3. The so-called "tailing parameter" $E_0$ phenomenologically describes the sub band-gap transitions or density of states, which are due either to interactions with phonons or to Coulomb potential fluctuations[17]. We finally remark that both PLE spectra show a dip at 6.15 eV, which may set the lower limit for the direct band gap of hBN [16].

The luminescence decay curves at different energies $E_{lum}$ are presented in Figure 4. Three luminescence bands at 5.47 eV, 5.56 eV and 5.61 eV show a similar quasi-mono-exponential (95%) decay with a characteristic time of τ = 2.5 ns. The shape of the 5.3-eV band decay is different: it is clearly multi-exponential and becomes slower with a decrease of the probed energy $E_{lum}$. This is confirmed by the time-gated luminescence spectra of the pBN sample presented in Figure 5a. In this sample, only the 5.3 eV band is observed. We have used two fast and medium time gates as defined in the part II. The 5.3-eV luminescence band exhibits a red shift when the time gate is delayed with respect to the SR excitation pulse. In other words, this result shows that the higher-energy part of the luminescence spectrum decays faster than the lower-energy one.

## IV. Discussion.

The time and energy-resolved photoluminescence experiments results clarify the nature of different luminescence bands. In the following, we assign three sub-bands at 5.5 eV to radiative transitions of bound excitons while the band at 5.3 eV to those of so-called quasi donor-acceptor pair (q-DAP). Additional supports to these assignments are given below.

*1. Bound excitons emissions lines*

We first discuss three sub-bands at 5.47, 5.56 and 5.61 eV observed in the powder-like pd-hBN sample. Their energy positions agree with those reported in cathodoluminescence by Silly *et al.*[3] (hBN powders) and by Watanabe *et al.*[1] (hBN single crystal). However, their relative intensities are different: in particular, the higher-energy component at 5.61 eV is the most intense in our spectra. Recently, these emissions have been assigned to bound-exciton luminescence caused by disorders such as stacking faults or shearing of the lattice planes [4].

Our photoluminescence excitation spectra support this assignment. Three 5.5-eV sub-bands appear at $E_{exc}$ > 5.70 eV (see Figure 2) and reach the maximum intensity at $E_{exc} = 5.8\ eV$ (see Figure 3). This energy corresponds to free exciton in hBN single cristal[1] (5.76 eV) and is predicted theoretically[14] (5.85 eV). The large width of the peak is explained by the contribution of four non-resolved transitions [14]. The PL spectra suggest that free excitons migrate and a part of them radiatively recombine on defects. The question is whether these sub-bands belong to an exciton bound to different defects (heterogeneous line-shape) or to a unique defect (homogeneous line-shape). Similar shapes of the luminescence decay curves make the first possibility more likely. Future studies of the temperature effect on fluorescence spectra may provide more information concerning this issue.

We remark that in our samples the free exciton luminescence at 5.76 eV (215 nm) is not observed. This may be related to a poor sample quality. Watanabe *et al.*[4] have shown that a mere pressure applied between two fingers can easily deform hBN because of the softness inherent to the layered crystalline structure. This deformation results in a dramatic change of the band-edge luminescence, which evolves from that of free exciton in a high quality stress-free crystal to that of bound exciton at 227 nm (5.46 eV) in a deformed crystal. Recent cathodoluminescence imaging of isolated micro-crystallites shows that the free exciton in hBN efficiently decays on lattice dislocations [18]. The press-machine used to prepare our rigid pallets from hBN powder inevitably induces such defects.

*2. Donor acceptor emission at 5.3 eV.*

The 5.3 eV band has been assigned earlier to bound excitons of Wannier[1] or Frenkel type[3]. We disagree with these interpretations. Actually, the red shift observed in the time-gated luminescence experiment and the multi-exponential decay are characteristic of a DAP transition:

$$A^o + D^o \rightarrow A^- + D^+ + h\upsilon_{DAP} \quad (1)$$

Indeed, the energy of the photons resulting from the radiative recombination of a DAP depends on the distance $R_{AD}$ between the donor and the acceptor and is given by the following expression, provided that the donor and the acceptor are not too close [17].

$$h\upsilon_{DAP} = E_g - E_D - E_A + \frac{e^2}{\varepsilon R_{AD}} \quad (2)$$

$E_g$ is the energy band gap of hBN, $E_A$ and $E_D$ are the acceptor and donor energies relative to the VB and CB edges and $\varepsilon$ is the dielectric constant. The transition probability generally decreases with an increase of $R_{AD}$. The recombination at low photon energy (large $R_{AD}$) is then slower than at high energy (small $R_{AD}$). Together with equation (2), this explains the red-shift of the luminescence band recorded with a delay with respect to the excitation SR pulse. As a result, we assign the 5.3-eV luminescence to electronic transitions between filled acceptor and donor states.

The characteristics of the DAP luminescence strongly depends on a way in which the DAP is populated. Three main excitation channels are listed below:

$$A^- + D^+ + h\upsilon_{exc} \rightarrow A^- + D^+ + e^- + h^+ \rightarrow A^o + D^o \quad (3)$$

$$A^- + D^+ + h\upsilon_{exc} \rightarrow A^o + D^+ + e^- \rightarrow A^o + D^o$$
$$A^- + D^+ + h\upsilon_{exc} \rightarrow A^- + D^o + h^+ \rightarrow A^o + D^o \quad (4)$$

$$A^- + D^+ + h\upsilon_{exc} \rightarrow A^o + D^o \quad (5)$$

According to channel 3, free CB electrons and VB holes are created and separately localized in the material. The luminescence, which includes contributions from different DAP, is broad and its energetic position is independent on the excitation energy ($E_{lum}$ = const) as typical in photoluminescence. The same is true when either the donor or the acceptor is directly ionised (channel 4), since the free carriers can be arbitrary trapped in the material. On the other hand, under condition of selective excitation (channel 5), a particular subset of DAP is populated which emission generally corresponds to the excitation energy. This luminescence may also be shifted due to internal relaxation within the low-lying DAP states or due to emission of phonons. In both cases, the shift between the excitation and luminescence energies is constant: $E_{exc} - E_{lum}$ = const. In that sense, selectively excited PL is similar to Raman scattering. Moreover, the luminescence is narrowed with respect to that resulting from the band-gap excitation since only a subset of DAP is excited. As a result, one can distinguish the PL (3,4) and the Raman (5) regimes of radiative transitions involving DAP levels.

As we mentioned in the part III, the 5.3 eV PLE spectra intensity in the onset region exponentially increases with $E_{exc}$ (Figure 3). This dependence can be understood if we assume that $I_{lum}$ is proportional to the absorption coefficient. Such an exponential dependence of the absorption coefficient is characteristic of an Urbach tail and conveys the density of states below the band gap. The "tailing parameter" $E_0$ = 41 meV obtained from the fit characterizes the spectral width of this tail of states. Its value of few tens of meV is generally reported in semiconductors with a strong local perturbations of the band structure by charged impurities[19]. This point will be helpful for the following discussion of the 5.3 eV band blue shift.

When the excitation energy is changed from 5.5 eV to 6 eV, the 5.3 eV band undergoes a blue shift. This blue energy-shift is rather large ~ 120 meV and characterized by two regimes represented in Figure 5b. (i). For excitation above 5.7 eV, the band position is fixed at 5.3 eV as expected for excitation conditions corresponding to channels (3-4). (ii) For excitation below 5.7 eV the band position $E_{lum}$ varies linearly with excitation energy $E_{exc}$. In the case of selective DAP excitation, such a dependence is expected with a slope $\Delta E_{lum} / \Delta E_{exc} = 1$. This is shown by a doted line in Figure 5b. In contrast, the linear fit of the experimental data results in a slope $\Delta E_{lum} / \Delta E_{exc} = 0.66$, which indicates that the energy difference $E_{exc}$ - $E_{lum}$ is not constant. This variable shift between $E_{lum}$ and $E_{exc}$ shows that a majority of the excited charge carriers do not annihilate on the initially excited DAPs. They efficiently relax via hoping into a low-energy DAP. The observed luminescence therefore involves DAP states different from those initially excited. This conclusion is supported by the constant spectral width (~180 meV) of the emission band, whatever the excitation energy is above or below the band gap. In the opposite case of localized excited carriers, the selective excitation of a particular subset of DAP should result into a narrowing of the DAP luminescence band as compared to the above band-gap excitation.

We assign the observed blue-shift of the 5.3 eV band to an increased number density of the photo-induced charge carriers. It is known that DAP transitions are subjected to a blue-shift as more charges are excited per unit volume. As a consequence, the donors and acceptors are forced to be closer to each other and the Coulomb term in equation (1) causes an increase of the transition energy. However, when the number of carrier is sufficiently high, no close pairs can be formed and the position of the luminescence band becomes fixed. In our experiment, we have observed this spectral crossover for excitation at 5.75 eV. However, the classical DAP recombination model cannot accommodate the linear dependence between excitation photon energy $E_{exc}$ and luminescence energy $E_{lum}$ with a slope <1, as well as the strong blue-shift observed in the present experiments. This probably suggests a recombination process involving Coulomb bands fluctuations. We discuss below this issue.

In highly-compensated materials, most of defects are ionized and the number density of free carriers is low. A random distribution of charged defects induces spatial fluctuations of the band structure that result into a broadening of the defect levels and the formation of band tails. The average amplitude $\Gamma$ of the potential fluctuations can be expressed as [20, 21]:

$$\Gamma = \frac{e^2}{\varepsilon} \cdot \frac{N_t^{2/3}}{g^{1/3}} \qquad (6)$$

where $N_t$ stands for the total concentration of charged defects ($N_t = N_D^+ + N_A^-$), and $g$ is the density of free charges according to the material type and excitation conditions. The other symbols have their usual meaning. The amplitude $\Gamma$ increases with the concentration of charged defects and decreases with that of free carriers. Radiative recombination in such materials is governed by the recombination of carriers localized in spatially separated potential minima originating from Coulomb potential fluctuations. This localization affects the energy of so-called quasi-DAP (q-DAP) transitions by twice the averaged fluctuation amplitude [20, 21]:

$$h\upsilon_{q-DAP} = h\upsilon_{DAP} - 2\Gamma \tag{7}$$

where $h\upsilon_{DAP}$ sets for the photon energy of DAP recombination in the flat-band case given by equation (2). Similar to recombination involving individual donors and acceptors, a shift towards higher photon energy is observed for q-DAP transitions. This effect is amplified by fluctuations of the electrostatic potential affected by the photo-excited charge number density. The last one depends on the excitation intensity $I_{exc}$: at low intensity donor states in the potential minima of the conduction band and acceptor states in the maxima of valence band will be preferably populated minimising the transition energy. When the intensity increases, the fluctuation amplitude $\Gamma$ decreases due to the neutralization of defect states by processes (3)-(4) and the increase of free-charge density. Large blue-shifts dependent on excitation intensity as $E_{lum} = A \cdot \log_{10}(I_{exc}) + C$ have been observed in many experiments[20, 22-25] and explained in the framework of the Coulomb potential fluctuation model. Assuming that the photo-excited charge density $N$ is proportional to excitation intensity $I_{exc}$, the observed energy shift can be rewritten as

$$E_{lum} = A \cdot \log_{10}(N) + C' \tag{8}$$

We apply this model to explain the 5.3-eV band shift (Figure 5b). The observation of an Urbach tail in the excitation spectrum supports the presence of the electrostatic potential fluctuations. We believe that the blue-shift observed in our experiment with increasing of excitation energy has the same origin as the one observed with an increase of the excitation intensity: in both cases, the photo-excited carrier density $N$ is affected. Because a permanent density of free charges is unlikely in our pulsed-excitation experiment, the spectral shift results more probably from the neutralization of the charged defects (reduction of $N_t$).

From the observed luminescence intensity in the Urbach tail $I_{lum} \propto \exp(E_{exc}/E_0)$ we can express the density of photo-excited charges as $N \propto \exp(E_{exc}/E_0)$. Combining it with Eq. (8), we obtain the following expression for the luminescence band position:

$$E_{lum} = \frac{A}{E_0 \cdot \ln(10)} E_{exc} + C'' \tag{9}$$

The least-square fit of the experimental points in Figure 5.b) by Eq. (9) (solid line) leads to an energy-shift constant A=62 meV, which is proportional to the band fluctuations amplitude $\Gamma$ [20]. This picture seems to be reasonable and is similar to the energy-shift constant that has been reported in Mg-doped GaN [22, 26].

As the PLE spectra in Figure 3 shows, the exciton peak at 5.8 eV contributes to the bound exciton emission at ~5.5 eV, while leaving the q-DAP emission at 5.3 eV unaffected. The last indicates that the exciton does not transfer its energy to the DAP involved in the luminescence process. This finding corroborates our band assignments. Actually, the exciton corresponding to bound charges cannot compensate charged defects unless it dissociates. In principle, this dissociation is possible only if the gain in energy, due to the charge localization, is greater than the exciton binding energy. This set the lower limit to exciton binding energy to $D_e > (E_g - E_{DAP})/2 \approx 0.4$ eV. This conclusion disagrees with the Wannier-type of hBN exciton suggested by Watanabe et al.[1] and supports Arnaud's prediction of Frenkel-type exciton possessing a strong binding energy[14]..

Finally, the identification of the nature of the DAP responsible for luminescence at 5.3 eV requires further experiments with high-quality hBN samples. Their elaboration and quantitative characterization of crystalline defects and impurities are delicate tasks. Nevertheless, we have recently shown that laser UV irradiation of hBN under vacuum promotes the creation of nitrogen vacancies and the sample enrichment by boron [11, 12, 27]. Future PL studies of these treated samples may be of interest.

## v.  Conclusion.

Near band gap luminescence (hν≥5 eV) of hexagonal boron nitride has been studied by time- and energy-resolved photoluminescence spectroscopy methods. Depending on the excitation energy, several luminescence bands have been observed in two kinds of samples made from compressed commercial powder (pd-hBN) and pyrolytic BN (pBN). Two principal emissions have been observed at 5.5 eV and 5.3 eV. High-resolution luminescence spectra of pd-hBN samples resolve three subbands of the high-energy emission at 5.47 eV, 5.56 eV and 5.61 eV, which are assigned to bound excitons. They appear at $E_{exc} \geq 5.7$ eV and are not observed in pyrolytic BN sample due to a lower quality. In contrast, the non-structured band at 5.3 eV with a multi-exponential decay is observed in both kinds of samples at $E_{exc} \geq 5.5$ eV. Its excitation spectrum shows the characteristic Urbach tail due to electrostatic band fluctuations induced by charged defects. We assign it to q-DAP radiative recombination. A striking feature of the 5.3-eV emission is its large blue shift (up to 120 meV) with a linear slope $\Delta E_{lum} / \Delta E_{exc}$ <1 when the excitation energy increases. At $E_{exc} \geq 5.7$ eV, the band position is fixed revealing the transition from the Raman to the PL regime. The shift can be explained in the framework of the band-fluctuation model by photo-induced neutralization of charged defect states. The absence of excitonic contribution to the DAP luminescence suggests a strong exciton binding energy, which is in agreement with recent theoretical predictions.[14]

# Acknowledgments

This work has been supported by the IHP-Contract HPRI-CT-1999-00040 of the European Commission. The authors are grateful to G. Stryganyuk for assistance in conducting experiments at SUPERLUMI station and to V. Solozhenko for helpful discussions and for kindly providing the pyrolytic boron nitride samples. Furthermore the authors thank J.P. Schermann and N. Braidi for reading the manuscript.

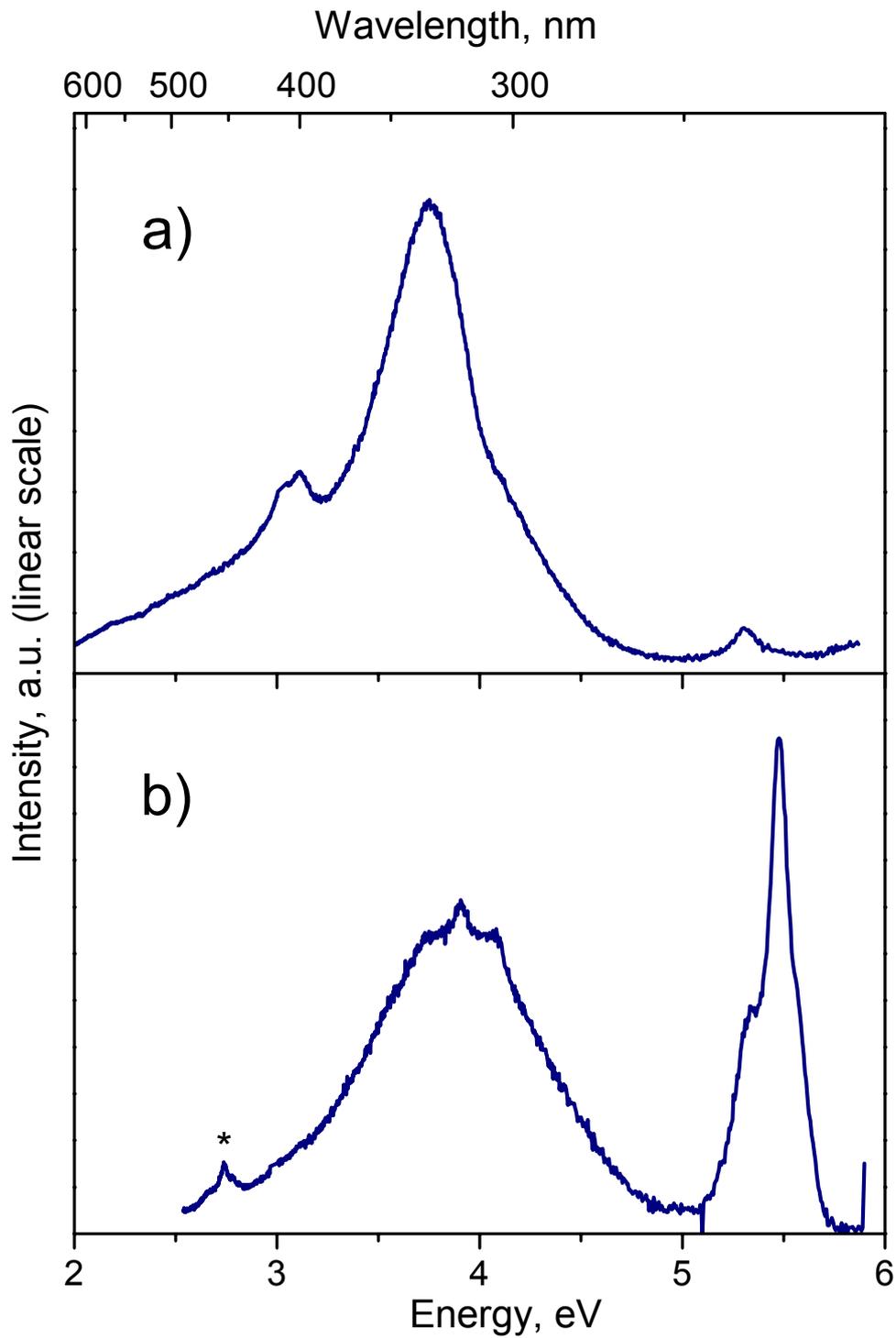

**Figure 1 :** Low-temperature luminescence spectra of pyrolitic BN (a) and powder-like hBN (b) excited with photons of 6.0 eV. The star in (b) indicates the second order of the 5.5 eV band.

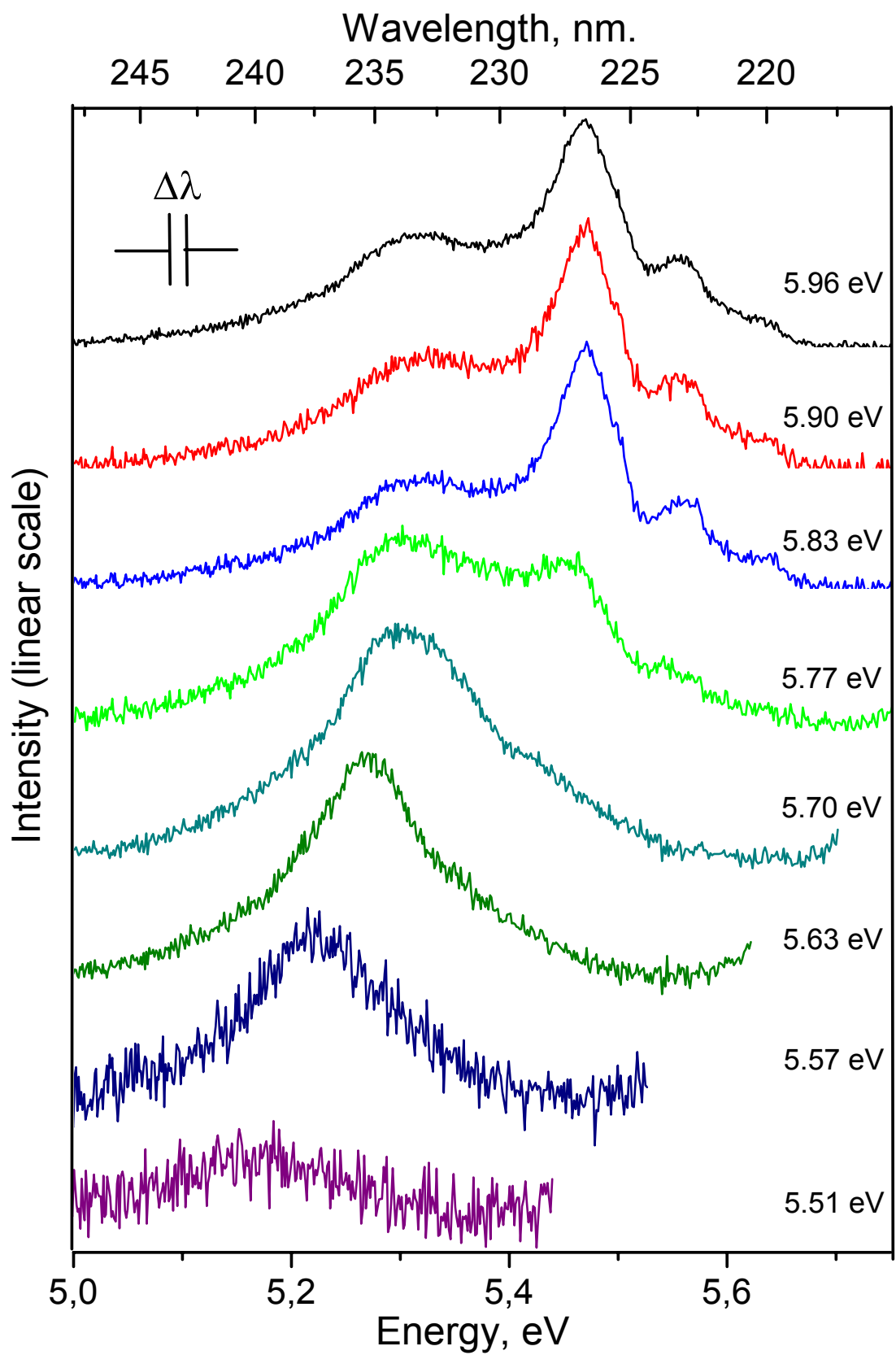

**Figure 2 :** Near band gap luminescence spectra of powder-like hBN (pd-hBN) for different SR-photons excitation energy.

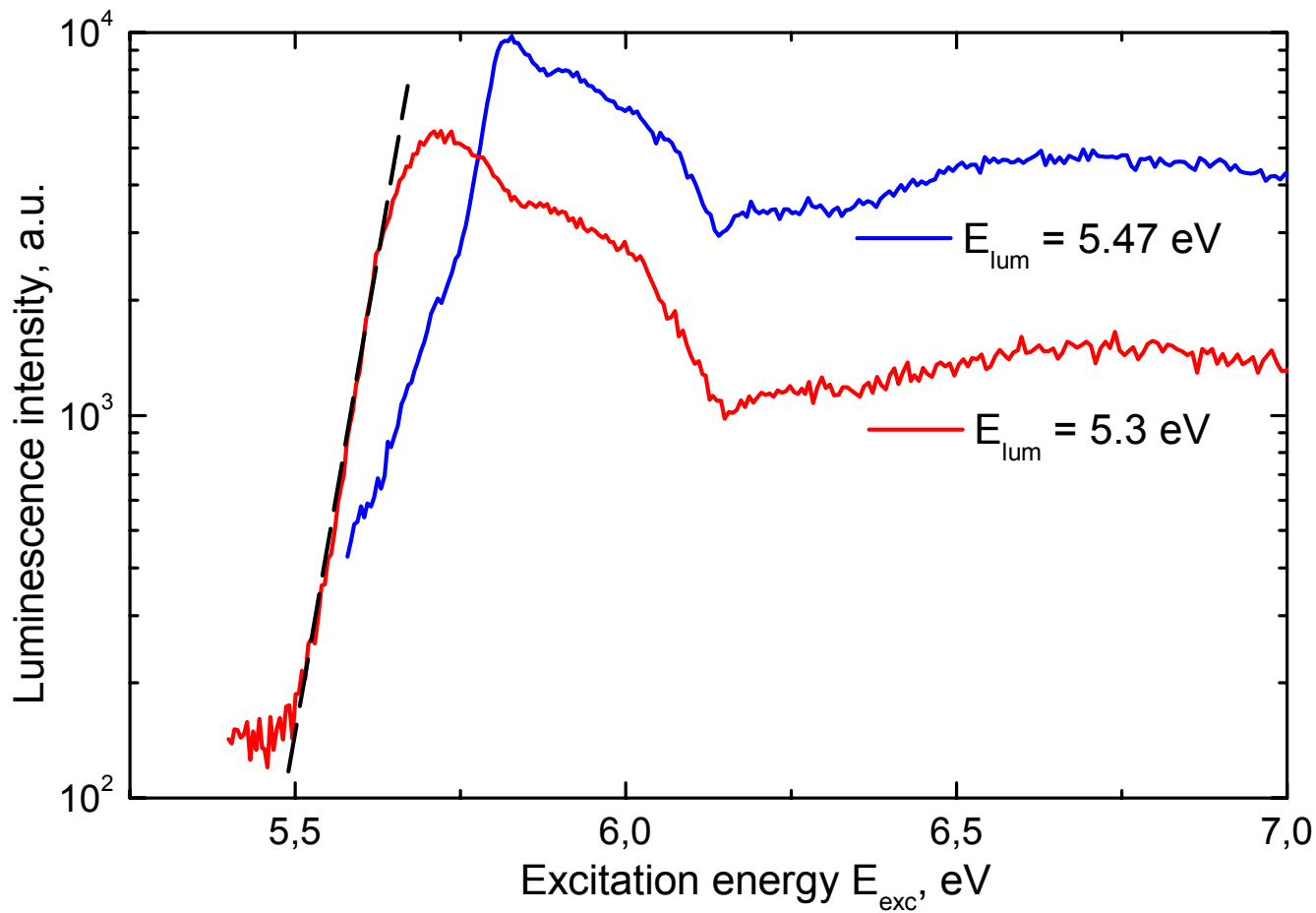

**Figure 3 :** Excitation spectra of hBN photoluminescence at 5.3 eV and 5.47 eV. The exponential fit of the low-energy part of the 5.3-eV band lineshape is shown by the dotted line (see in text).

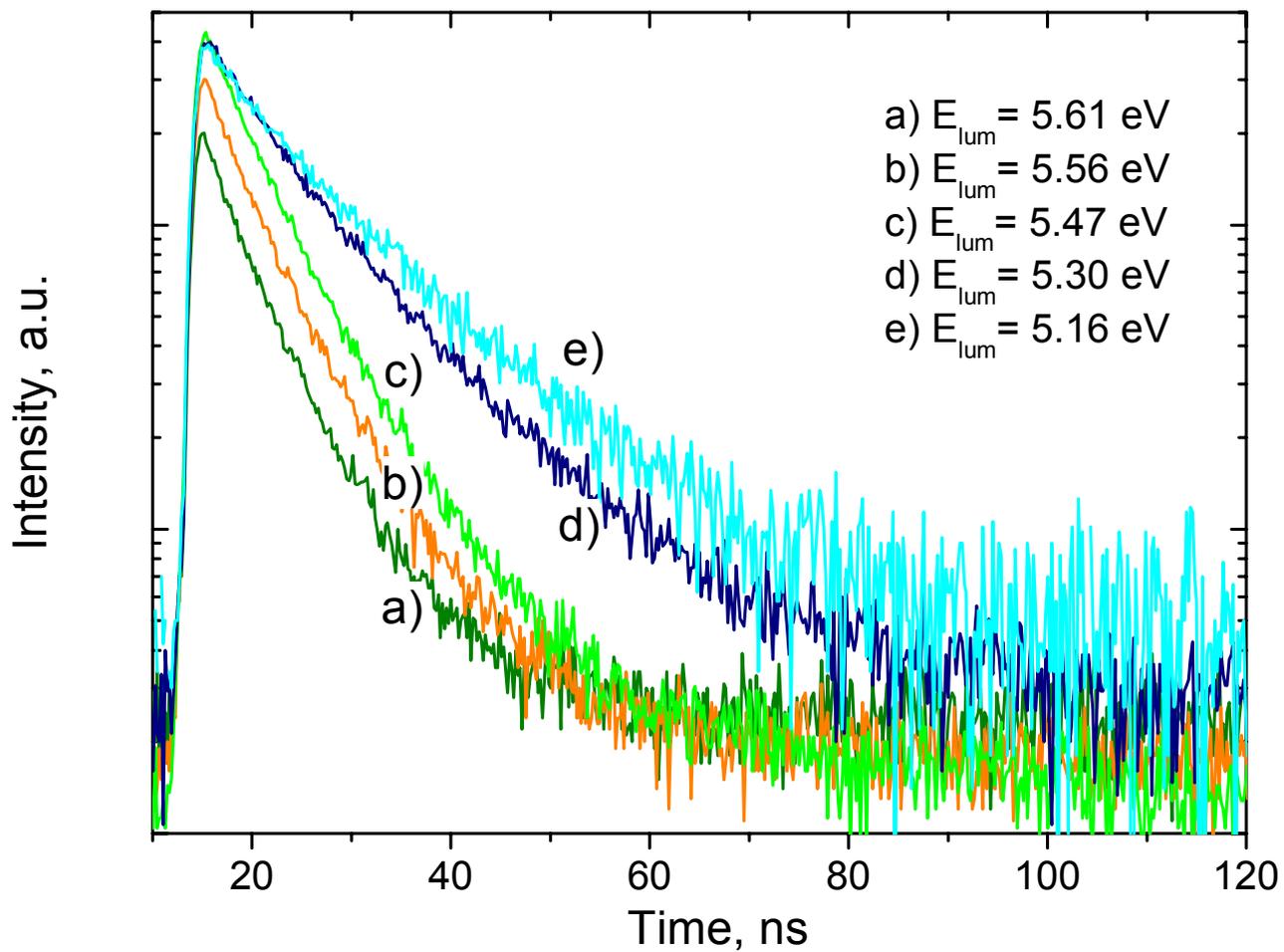

**Figure 4 :** Luminescence decay curves of powder-like hBN excited at 6.0 eV and probed at energies indicated on the figure. The intensity scale is not common for all curves.

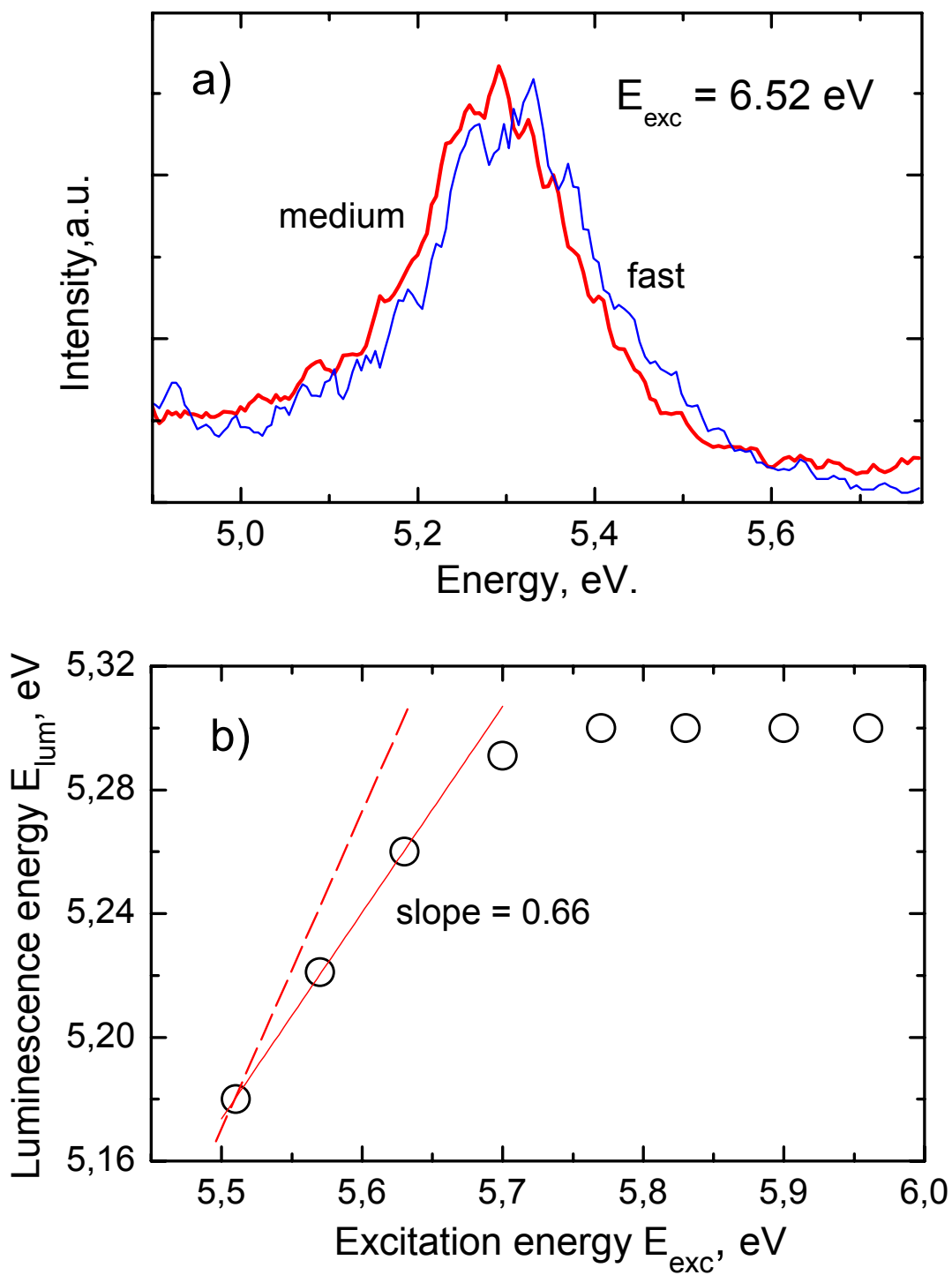

**Figure 5 :** Luminescence spectra of pBN recorded with fast (Δτ=2-6 ns) and medium time frames (Δτ=10-30 ns) (a). Spectral position of the pd-hBN luminescence band versus the excitation photon energy (b). The solid and dashed lines represent the linear least square fit with a slope equal to 0.66 and the expected variation in the case of the selective excitation with a slope equal to 1, respectively.